\documentclass[11pt,a4paper]{article}
\usepackage[margin=1in]{geometry}
\usepackage[T1]{fontenc}
\usepackage[utf8]{inputenc}
\usepackage{lmodern}
\usepackage{amsmath,amssymb,amsthm,mathtools}
\usepackage{microtype}
\usepackage[numbers]{natbib}
\usepackage{xurl}
\usepackage[hidelinks]{hyperref}
\usepackage{bookmark}
\theoremstyle{plain}
\newtheorem{theorem}{Theorem}
\newtheorem{lemma}[theorem]{Lemma}
\newtheorem{corollary}[theorem]{Corollary}

\theoremstyle{definition}
\newtheorem{definition}[theorem]{Definition}
\theoremstyle{remark}

\newcommand{\HH}{\mathcal H}
\newcommand{\FF}{\mathcal F}
\newcommand{\mathcalA}{\mathcal A}
\newcommand{\BB}{\mathcal B}
\newcommand{\QQ}{\mathcal Q}
\newcommand{\RR}{\mathcal R}
\newcommand{\vc}{\operatorname{VCdim}}

\newcommand{\E}{\mathbb E}
\newcommand{\Prb}{\mathbb P}
\newcommand{\eps}{\varepsilon}

\title{Improved Bounds on Polychromatic Colorings of $t$-tuples}
\date{}
\hypersetup{
  pdftitle={Improved Bounds on Polychromatic Colorings of t-tuples},
  pdfsubject={Polychromatic colorings of tuples in geometric and bounded-VC hypergraphs},
  pdfkeywords={Polychromatic coloring, tuple coloring, pseudo-discs, VC-dimension, epsilon-nets, lower shadows}
}
\author{Shakhar Smorodinsky\thanks{%
Institute for the Theory of Computing,
Stein Faculty of Computer and Information Science,
Ben-Gurion University of the Negev, Beer-Sheva, Israel.
Email: \href{mailto:shakhar@bgu.ac.il}{\texttt{shakhar@bgu.ac.il}}.}}
\hypersetup{pdfauthor={Shakhar Smorodinsky}}

\begin{document}
\maketitle
\begin{abstract}
We prove that there is a function $f(k)=O(k)$ such that the pairs
of points in any finite point set $P$ in the plane can be $k$-colored
so that any disc containing at least $f(k)$ points of $P$ contains
a pair of each color. This improves the previous exponential upper
bound of $3.7^k$ due to Biniaz et al.\ (SoCG '25) and answers
their question of whether a polynomial bound is possible.
Our linear bound also holds for families of pseudo-discs.

More generally, let $H=(V,E)$ be a hypergraph. We seek a $k$-coloring
of the $t$-element subsets of $V$ such that every hyperedge with
at least $f$ vertices contains a tuple of each color.
For hypergraphs defined by points and pseudo-discs in the plane,
we obtain $f=O_t(k^{1/(t-1)})$ for every fixed $t\geq 2$.
For shrinkable hypergraphs of VC-dimension at most $d$, we prove
that $f=O_{d,t}(k^{1/(t-d)})$ for every fixed $t>d$, improving
in particular the previous exponential bound for $(d+1)$-tuples
to a linear one.

Our method combines random sampling, bounds on the lower shadows
of families of shallow tuples, a capacitated version of Hall's
theorem, and the $\eps$-net theorem. The key idea is to assign
shallow tuples to smaller subsets, thereby reducing the global
coloring problem to a collection of local coloring problems
in which the relevant hyperedges contain a constant fraction
of the ground set.
\end{abstract}

\section{Introduction}
A hypergraph $H$ is a pair $H=(V,E)$, where $V$ is a finite set and $E$ is a family of subsets of $V$. The elements of $V$ are called \emph{vertices} and those of $E$ are called \emph{hyperedges}. A \emph{$t$-tuple} is simply an unordered $t$-element subset of $V$ (not necessarily in $E$). A coloring $\chi:\binom Vt\to[k]$, where $[k]=\{1,\ldots,k\}$, is said to be $(t,k,f)$-polychromatic if every hyperedge of size at least $f$ contains a tuple of each of the $k$ colors. Let $f_H(t,k)$ denote the minimum integer $f$ for which $H$ admits a $(t,k,f)$-polychromatic coloring. For a family $\HH$ of hypergraphs, let $f_{\HH}(t,k)=\sup_{H\in\HH}f_H(t,k)$.

A concrete example is the family $\HH_{\mathrm{disc}}$ of hypergraphs whose vertices are the points of a finite planar point set $P$ and whose hyperedges are the subsets of $P$ that can be cut out by discs. More generally, for a finite planar point set $P$ and a family of regions $\RR$, write $H(P,\RR)=(P,\{P\cap R:R\in\RR\})$.

Polychromatic coloring of vertices ($t=1$) provides a way to partition a ground set into disjoint hitting sets for all sufficiently large hyperedges. In particular, a $(1,k,f)$-polychromatic coloring yields $k$ pairwise disjoint $\eps$-nets whenever $f\leq\lceil\eps|V|\rceil$. The study of geometric polychromatic colorings goes back to the work of Pach in 1980~\cite{Pach80} on cover-decomposition: can a sufficiently thick covering by translates of a fixed planar convex body be split into two coverings? This is a polychromatic coloring problem in the dual geometric hypergraph, whose vertices are the covering regions and whose hyperedges are the subfamilies of regions containing a given point. Apart from its intrinsic geometric interest, this problem has applications to sensor networks, where disjoint groups of sensors can be activated in succession, each maintaining coverage of the targets~\cite{PPTsurvey}. Among the major positive results are the optimal threshold $2k-1$ for halfplanes~\cite{SY} and the linear threshold for translates of any fixed convex polygon~\cite{GV}. Polynomial bounds were also obtained for translates of an octant in $\mathbb R^3$, and hence for homothets of a triangle in the plane~\cite{CKMU}. On the other hand, Chen et al.~\cite{CPST} showed in 2009 that no finite threshold exists for coloring points with two colors with respect to axis-parallel rectangles. In 2016, Pach and P\'alv\"olgyi~\cite{PP} answered the original question for unit discs in the negative, showing that no finite bichromatic threshold exists even in this setting. The study of polychromatic colorings has also led to connections with discrepancy theory and linear programming in abstract hypergraphs~\cite{BPRS}. For recent developments in geometric settings, see the work on shallow hitting sets~\cite{PU} and the survey~\cite{DKPPT}.

The above negative results motivate the study of coloring tuples of vertices rather than individual vertices. For tuple colorings ($t\geq2$), Ackerman, Keszegh, and P\'alv\"olgyi~\cite{AKP} proved polynomial bounds for axis-parallel boxes in any fixed dimension and studied the more restrictive problem of coloring Delaunay hyperedges.

Biniaz, De Carufel, Maheshwari, Smid, Smorodinsky, and Stojakovi\'c~\cite{BDMSSS,BDMSSSfull} proved that $f_{\HH_{\mathrm{disc}}}(2,k)\leq 3.7^k$ for discs and obtained an upper bound of $c^k$ for pseudo-discs, and explicitly asked whether a polynomial bound is possible. They also proved exponential bounds for $(d+1)$-tuples in shrinkable hypergraphs of VC-dimension at most $d$, using the existence of tuples of large depth. We answer their question affirmatively and obtain polynomial bounds in all these settings using a different method.

\paragraph{Our results.}
A family of pseudo-discs is a collection of closed Jordan regions in the plane such that the boundaries of any two distinct regions intersect at most twice. Let $\HH_{\mathrm{pd}}$ consist of the hypergraphs induced by arbitrary finite point sets and arbitrary families of pseudo-discs in the plane. In particular, $\HH_{\mathrm{pd}}$ includes the hypergraphs induced by discs.

\begin{theorem}\label{thm:pseudodisks}
For every fixed integer $t\geq 2$ there exists a constant $C_t$ such that, for every $k\geq 2$,
\[
 f_{\HH_{\mathrm{pd}}}(t,k)\leq \left\lceil C_t k^{1/(t-1)}\right\rceil.
\]
In particular, $f_{\HH_{\mathrm{pd}}}(2,k)=O(k)$.
\end{theorem}

This substantially improves the previous exponential bound for pseudo-discs. 

A hypergraph is said to be \emph{shrinkable} if for every $e\in E$ and every integer $1\leq j\leq |e|$ there exists $e'\in E$ with $e'\subseteq e$ and $|e'|=j$. For the proof at a given threshold $m$, we will need only the weaker assumption that every hyperedge of size at least $m$ contains a hyperedge of size exactly $m$. Denote by $\HH_d$ the family of all finite shrinkable hypergraphs of VC-dimension at most $d$.

\begin{theorem}\label{thm:vc}
For fixed integers $d\geq 1$ and $t>d$ there exists a constant $C_{d,t}$ such that, for every $k\geq 2$,
\[
 f_{\HH_d}(t,k)\leq \left\lceil C_{d,t} k^{1/(t-d)}\right\rceil.
\]
In particular, $f_{\HH_d}(d+1,k)=O_d(k)$.
\end{theorem}

For a family with arbitrarily large shrinkable hyperedges, the inequality $\binom ft\geq k$ gives the counting lower bound $f_{\HH}(t,k)=\Omega_t(k^{1/t})$. Our results replace the previous exponential dependence on $k$ by polynomial dependence, with constants independent of the ground-set size. In particular, for pairs of points and discs, $f_{\HH_{\mathrm{disc}}}(2,k)$ is between $\Omega(\sqrt{k})$ and $O(k)$.

\paragraph{Proof outline.}

To illustrate the main ideas of our approach, we first sketch the proof of
the linear bound for coloring pairs of points with respect to discs in
the plane. The proof combines three ingredients: the sparsity of the
graph of shallow pairs, a degeneracy ordering of this graph, and the
existence of constant-size $\varepsilon$-nets for hypergraphs of bounded
VC-dimension.

Let $P$ be a finite set of points in the plane and let $m$ be an integer
to be determined later. Define the graph $G_m$ on $P$, where two points
$p,q\in P$ are adjacent if there exists a disc containing both $p$ and $q$
and at most $m$ points of $P$. The first step is to observe that $G_m$ is
$O(m)$-degenerate. This follows from a random sampling argument showing that every induced subgraph $G_m[X]$ on a subset $X\subseteq P$ has $O(m|X|)$ edges.

Consider a degeneracy ordering $p_1,\ldots,p_n$ of the vertices of $G_m$,
and let $N^+(p)$ denote the set of neighbors of $p$ that appear after $p$
in this ordering. Thus, $|N^+(p)|\leq am$ for some absolute constant $a$.
We assign each edge $pq$ of $G_m$ to its earlier endpoint in the ordering, say $p$.
The idea is to color the pairs in groups: for each $p$, the pairs $pq$ with $q\in N^+(p)$ are colored independently of the other pairs. Pairs that are not edges of $G_m$ are colored arbitrarily.

Now consider any disc $D$ containing exactly $m$ points of $P$, and let
$p$ be the first point of $D\cap P$ in the degeneracy ordering. Since
$D$ itself witnesses that every pair of points in $D\cap P$ is an edge
of $G_m$, we have
\[
    (D\cap P)\setminus\{p\}\subseteq N^+(p).
\]
In particular, $D$ contains a constant fraction of the points of
$N^+(p)$.

The problem thus reduces to coloring each set $N^+(p)$ independently
with $k$ colors such that every disc containing at least $m-1$ of its
points contains points of all $k$ colors. Using the $\varepsilon$-net
theorem, one can obtain such a coloring whenever $m\geq Ck$ for a
sufficiently large absolute constant $C$.

We color each edge $pq$ of $G_m$, where $p$ precedes $q$ in
the ordering, with the color assigned to $q$ in the coloring of
$N^+(p)$. Since every edge is assigned to a unique vertex, these local
colorings define a consistent global coloring. Every disc containing
exactly $m$ points contains pairs of all $k$ colors, all incident to
its earliest point in the ordering. 
Since every disc containing at least $m$ points of $P$ has a disc trace of size exactly $m$ contained in its trace, the same holds for every such disc.

The main idea of our proof is therefore to localize the coloring
problem by assigning shallow pairs to individual vertices. For larger
tuples and pseudo-discs, we use the same ordering and obtain tuple
colors from combinations of vertex colors in each set $N^+(p)$.
For arbitrary shrinkable hypergraphs of bounded VC-dimension, we extend
the assignment step by assigning each shallow tuple to one of its
smaller subsets, using bounds on lower shadows and Hall's marriage theorem.

We first give the full proof for pairs and discs in Section~\ref{sec:discs}. In Section~\ref{sec:disks} we extend it to pseudo-discs and to larger tuples. Finally, in Section~\ref{sec:vc}, we show how the assignment of tuples to smaller subsets replaces the degeneracy ordering for arbitrary shrinkable hypergraphs of bounded VC-dimension.

\section{Coloring pairs of points with respect to discs in the plane}
\label{sec:discs}
We prove that $f_{\HH_{\mathrm{disc}}}(2,k)=O(k)$. Fix a finite point set $P$ in the plane and an integer $m\geq2$. We call a pair of points of $P$ \emph{$m$-shallow} if there exists a disc containing both points and at most $m$ points of $P$. Let $G_m$ be the graph on $P$ whose edges are the $m$-shallow pairs.

\begin{lemma}\label{lem:degenerate}
The graph $G_m$ is $\lfloor6\mathrm e\,m\rfloor$-degenerate, where $\mathrm e$ is the base of the natural logarithm.
\end{lemma}
\begin{proof}
We use a standard random sampling argument (in the spirit of Clarkson and Shor~\cite{CS}). Let $X\subseteq P$. For every edge $pq$ of $G_m[X]$, fix a disc $D_{pq}$ that contains $p,q$ and at most $m$ points of $P$. Let $Y\subseteq X$ be a random sample obtained by choosing each point of $X$ independently with probability $p_0=1/m$. With probability at least
\[
 p_0^2(1-p_0)^{m-2}\geq\mathrm e^{-1}p_0^2,
\]
we have $D_{pq}\cap Y=\{p,q\}$. In that event $pq$ is an edge of the Delaunay graph of $Y$, namely, the graph whose edges are the pairs that can be cut out by a disc. This graph is planar and hence has at most $3|Y|$ edges. By linearity of expectation,
\[
 \mathrm e^{-1}m^{-2}|E(G_m[X])|\leq3\E|Y|=3|X|/m.
\]
Thus $|E(G_m[X])|\leq3\mathrm e\,m|X|$. Every nonempty induced subgraph of $G_m$ therefore has a vertex of degree at most $\lfloor6\mathrm e\,m\rfloor$, proving the lemma.
\end{proof}

Take a degeneracy ordering $p_1,\ldots,p_n$ of the vertices of $G_m$. For each $p\in P$, let $N^+(p)$ be the set of neighbors of $p$ that appear after $p$ in the ordering. Thus
\[
 |N^+(p)|\leq6\mathrm e\,m.
\]
We assign each $m$-shallow pair $pq$ to its earlier endpoint, say $p$. To color these pairs, we will independently color the points in each set $N^+(p)$. The following simple lemma will be useful here and in the more general settings considered later.

For a hypergraph $H=(V,E)$ and $X\subseteq V$, let $H[X]=(X,\{e\cap X:e\in E\})$ be the trace of $H$ on $X$. We say that $H$ shatters $X$ if $\{e\cap X:e\in E\}=2^X$. The VC-dimension $\vc(H)$ of $H$ is the largest size of a shattered set. Recall that passing to a trace does not increase VC-dimension. An $\eps$-net for $H$ is a subset of $V$ intersecting every nonempty hyperedge containing at least $\eps|V|$ vertices. By the $\eps$-net theorem of Haussler and Welzl~\cite{HW}, for fixed $D$ and $\eps>0$, there is a constant $b=b(D,\eps)$ such that every finite hypergraph of VC-dimension at most $D$ has an $\eps$-net of size at most $b$.

\begin{lemma}\label{lem:local}
Fix $D\geq 1$ and constants $0<B\leq A$. There is a constant $c=c(D,A,B)>0$ with the following property. Suppose $H=(W,\FF)$ has VC-dimension at most $D$ and $|W|\leq A M$ for a real number $M\geq1$. If $k\leq cM$, then $W$ admits a $k$-coloring such that every hyperedge of cardinality at least $BM$ contains all $k$ colors.
\end{lemma}
\begin{proof}
Set $\eps=B/(2A)$ and let $b=b(D,\eps)\geq1$ be as in the $\eps$-net theorem. Take $c=B/(2b)$. If there are no hyperedges of size at least $BM$, color arbitrarily. Otherwise choose an $\eps$-net $N_1$ for $H$, assign its vertices color 1, and remove them. Continue by choosing an $\eps$-net $N_i$ in the trace on the remaining ground set and assigning its vertices color $i$. Each net has size at most $b$; before the $i$th step, at most $(i-1)b\leq kb\leq BM/2$ vertices have been removed. Thus each original hyperedge of size at least $BM$ still has at least $BM/2$ vertices, which is at least an $\eps$-fraction of the remaining ground set. The $i$th net therefore hits each such hyperedge. After $k$ steps, color any remaining vertices arbitrarily.
\end{proof}

\paragraph{Coloring the pairs.}
For each $p\in P$, consider the hypergraph on $N^+(p)$ whose hyperedges are the subsets cut out by discs. Its VC-dimension is at most $3$. Applying Lemma~\ref{lem:local} with $M=m$, $A=6\mathrm e$, and $B=1/2$, we can $k$-color the points of $N^+(p)$ so that every disc containing at least $m-1$ points of $N^+(p)$ contains points of all $k$ colors, provided that $k\leq cm$ for some absolute constant $c>0$. Here we use $m-1\geq m/2$.

For every edge $pq$ of $G_m$, where $p$ precedes $q$, assign to $pq$ the color of $q$ in the coloring of $N^+(p)$. Color all other pairs arbitrarily. Since each edge is assigned to exactly one point, the colorings of the different sets $N^+(p)$ do not interfere with each other.

Let $D$ be any disc containing exactly $m$ points of $P$, and let $p$ be the first point of $D\cap P$ in the degeneracy ordering. For every other point $q\in D\cap P$, the disc $D$ witnesses that $pq$ is an edge of $G_m$, and $q$ appears after $p$. Hence
\[
 (D\cap P)\setminus\{p\}\subseteq N^+(p).
\]
In particular, $D$ contains $m-1$ points of $N^+(p)$, so it contains a point $q_i\in N^+(p)$ of each color $i\in[k]$. The pairs $pq_i$ give all $k$ colors in $D$.

Finally, any disc containing at least $m$ points of $P$ has a disc trace of size exactly $m$ contained in its trace. Indeed, slightly enlarge the original disc so that its trace is unchanged and all its points lie in the interior. Then perturb its center slightly, preserving the trace and making all distances to points of $P$ distinct. Shrink the resulting disc about its center until exactly $m$ points remain. Consequently, the coloring is polychromatic on every disc containing at least $m$ points. Taking $m=\lceil Ck\rceil$ for a sufficiently large absolute constant $C$ proves the claimed linear bound.

\section{From discs to pseudo-discs and larger tuples}
\label{sec:disks}
The proof in Section~\ref{sec:discs} uses three properties of discs: planarity of their Delaunay graphs, bounded VC-dimension, and shrinkability. We explain how to use these properties for pseudo-discs, and then extend the argument from pairs to $t$-tuples.

\paragraph{Coloring pairs with respect to pseudo-discs.}
Let $P$ be a finite planar point set and $\RR$ a family of pseudo-discs. By retaining one representative of each distinct trace on $P$, we may assume that $\RR$ is finite. The Delaunay graph on any finite point set with respect to a family of pseudo-discs, whose edges are the two-element traces, is planar~\cite{BPR,KP}. Therefore the sampling proof of Lemma~\ref{lem:degenerate} applies unchanged to the graph $G_m$ in which $pq$ is an edge if some pseudo-disc of $\RR$ contains $p,q$ and at most $m$ points of $P$. Also, the hypergraph $H(P,\RR)$ has VC-dimension at most $4$: if five points were shattered, their Delaunay graph would be $K_5$, contradicting planarity. Thus the coloring argument for pairs extends to the hyperedges of size exactly $m$, with $D=4$ in Lemma~\ref{lem:local}.

Unlike the family of all discs, an arbitrary pseudo-disc family need not be shrinkable. However, by a shrinking lemma of Pinchasi~\cite{Pinchasi} (see Lemma 18 in~\cite{KP}), if a pseudo-disc $R$ of a finite family contains more than $m\geq2$ points of $P$, there exists a pseudo-disc $R'\subset R$ containing exactly $m$ points of $P$ that can be added to the family while preserving the pseudo-disc property. Apply the lemma successively to each of the original pseudo-discs containing more than $m$ points, using the entire enlarged family at each step. We obtain a finite pseudo-disc family $\RR'$ containing $\RR$ such that every original hyperedge of size at least $m$ contains an $m$-element hyperedge of $H(P,\RR')$. Applying the preceding coloring argument to $\RR'$ gives the desired linear bound for $\RR$.

\paragraph{Coloring larger tuples.}
The extension from pairs to $t$-tuples is particularly simple. We use the observation of Biniaz et al.~\cite{BDMSSS,BDMSSSfull} that a polychromatic coloring of vertices with $\ell$ colors induces a polychromatic coloring of $r$-tuples with $\binom{\ell}{r}$ colors: the color of a tuple whose vertices have distinct colors is the set of its vertex colors.

\begin{proof}[Proof of Theorem~\ref{thm:pseudodisks}]
Fix an integer $m\geq2t$ and enlarge the pseudo-disc family $\RR$ to $\RR'$ as above. Define $G_m$ with respect to $\RR'$ and take a degeneracy ordering of its vertices. As before, let $N^+(p)$ denote the set of later neighbors of $p$, so that $|N^+(p)|\leq6\mathrm e\,m$ for every $p$.

Call a $t$-tuple \emph{$m$-shallow} if it is contained in a pseudo-disc of $\RR'$ containing at most $m$ points of $P$. The vertices of any such tuple induce a clique in $G_m$. Assign each $m$-shallow $t$-tuple $T$ to its first point $p$ in the ordering. Its remaining $t-1$ points belong to $N^+(p)$. We call $T\setminus\{p\}$ the \emph{completion} of $T$ at $p$.

As in the proof for pairs, by Lemma~\ref{lem:local} we can independently color the points in each $N^+(p)$ with $\ell=\lfloor cm\rfloor$ colors, for some absolute constant $c>0$, so that every pseudo-disc containing at least $m-1$ points of $N^+(p)$ contains all $\ell$ colors. Take $m$ large enough that $\ell\geq t-1$. We associate a distinct color with each $(t-1)$-element subset of $[\ell]$. Color an $m$-shallow $t$-tuple $T$ assigned to $p$ with the color corresponding to the set of vertex colors of its completion, if these are all distinct. Color all other tuples arbitrarily. Since each shallow tuple is assigned to a unique point, the coloring is well-defined.

Let $R\in\RR'$ contain exactly $m$ points of $P$ and let $p$ be the first of these points in the ordering. As before, $(R\cap P)\setminus\{p\}\subseteq N^+(p)$. Thus $R$ contains at least one point of each of the $\ell$ vertex colors in $N^+(p)$. For each $(t-1)$-element subset of $[\ell]$, choose one point of each of its colors in $R\cap N^+(p)$. These points, together with $p$, form an $m$-shallow $t$-tuple assigned to $p$ of the corresponding color. Hence $R$ contains tuples of all $\binom{\ell}{t-1}$ colors.

Since $\binom{\ell}{t-1}=\Theta_t(m^{t-1})$ for sufficiently large $m$, taking $m=\lceil C_t k^{1/(t-1)}\rceil$ with $C_t$ sufficiently large gives at least $k$ colors. Merge surplus colors to obtain exactly $k$ colors. By the choice of $\RR'$, the coloring is polychromatic on every original pseudo-disc containing at least $m$ points of $P$, as required.
\end{proof}

\begin{corollary}\label{cor:homothets}
Theorem~\ref{thm:pseudodisks} holds for the family of all homothets of any fixed planar convex body, with a constant depending only on $t$.
\end{corollary}
\begin{proof}
Homothets of a fixed strictly convex planar body form a pseudo-disc family (see~\cite{KP} for the convex setting). For a general convex body, choose one homothet realizing each of the finitely many traces on $P$. Fix an interior point of the body and slightly enlarge each chosen homothet about the corresponding interior point so that all included points lie in its interior and no excluded point is added. Approximate the original body sufficiently closely in the Hausdorff metric by a smooth strictly convex body. Because there are only finitely many chosen homothets and all point--boundary incidences have been made strict, their traces are unchanged under a sufficiently close approximation. Apply Theorem~\ref{thm:pseudodisks} to the approximating homothets; the constant is independent of the chosen body.
\end{proof}

\section{Shrinkable hypergraphs of bounded VC-dimension}
\label{sec:vc}
\label{sec:criterion}
In the geometric setting, the degeneracy ordering allowed us to assign every $m$-shallow $t$-tuple to one of its vertices, with at most $O_t(m^{t-1})$ tuples assigned to each vertex. In an arbitrary hypergraph of bounded VC-dimension, an analogous assignment to individual vertices need not exist. We therefore generalize the assignment step: each $m$-shallow $t$-tuple will be assigned to one of its $s$-element subsets, where $1\leq s<t$. Our goal is to ensure that at most $O(m^{t-s})$ tuples are assigned to any one $s$-subset. An averaging argument will then replace the choice of the first vertex in the degeneracy ordering.

\paragraph{A lower-shadow criterion.}
For integers $1\leq s<t$ and a family $\mathcalA\subseteq\binom Vt$, define its $s$-shadow by
\[
 \partial_s\mathcalA=\{S\in\tbinom Vs:S\subseteq T\text{ for some }T\in\mathcalA\}.
\]
For a hypergraph $H=(V,E)$ and an integer $m\geq t$, let
\[
 \mathcalA_m(H)=\{T\in\tbinom Vt:\exists e\in E\text{ with }T\subseteq e,\ |e|\leq m\}.
\]
As in the geometric setting, we call the members of $\mathcalA_m(H)$ the $m$-shallow $t$-tuples.

\begin{definition}\label{def:shadow}
For fixed $s<t$, a hypergraph $H$ satisfies the $(s,t,a)$-shadow condition if for every $X\subseteq V$ and every family $\BB\subseteq E(H[X])\cap\binom Xt$,
\begin{equation}\label{eq:exact-shadow}
 |\BB|\leq a|\partial_s\BB|.
\end{equation}
\end{definition}

The condition applies to \emph{arbitrary subfamilies} of the exact $t$-element hyperedges of every trace. A bound only on the total number of such hyperedges in terms of $|X|$ would not, in general, suffice for the matching argument below.

\begin{lemma}[Sampling and shadows]\label{lem:sampling}
Suppose $H$ satisfies the $(s,t,a)$-shadow condition. For every $m\geq t$ and every $\mathcalA\subseteq\mathcalA_m(H)$,
\begin{equation}\label{eq:shallow-shadow}
 |\mathcalA|\leq \mathrm e\,a\,m^{t-s}|\partial_s\mathcalA|,
\end{equation}
where $\mathrm e$ denotes the base of the natural logarithm.
\end{lemma}
\begin{proof}
For every $T\in\mathcalA$, fix a witnessing hyperedge $e_T$ containing $T$ with $|e_T|\leq m$. Take a random sample $X\subseteq V$ by retaining each vertex independently with probability $p=1/m$. Define
\[
 \mathcalA_X=\{T\in\mathcalA:e_T\cap X=T\}.
\]
Each member of $\mathcalA_X$ is an exact $t$-element hyperedge of $H[X]$. For every $T\in\mathcalA$,
\begin{align*}
 \Prb(T\in\mathcalA_X)
 &=p^t(1-p)^{|e_T|-t}\\
 &\geq p^t(1-p)^{m-t}\geq \mathrm e^{-1}p^t.
\end{align*}
In the last inequality, $t\geq1$ and $(1-1/m)^{m-1}\geq\mathrm e^{-1}$ for $m\geq2$. Thus $\E|\mathcalA_X|\geq\mathrm e^{-1}p^t|\mathcalA|$. On the other hand, by~\eqref{eq:exact-shadow},
\[
 |\mathcalA_X|\leq a|\partial_s\mathcalA_X|,
 \qquad \partial_s\mathcalA_X\subseteq (\partial_s\mathcalA)\cap\binom Xs.
\]
Consequently $\E|\mathcalA_X|\leq ap^s|\partial_s\mathcalA|$. Comparing the two inequalities and substituting $p=1/m$ proves~\eqref{eq:shallow-shadow}.
\end{proof}

\begin{lemma}[Assignment of shallow tuples]\label{lem:assignment}
Under the hypotheses of Lemma~\ref{lem:sampling}, there is a mapping $\alpha:\mathcalA_m(H)\to\binom Vs$ such that $\alpha(T)\subseteq T$ for every $T$ and
\[
 |\alpha^{-1}(S)|\leq L:=\left\lceil\mathrm e\,a\,m^{t-s}\right\rceil
 \qquad\text{for every }S\in\binom Vs.
\]
\end{lemma}
\begin{proof}
Construct a bipartite graph with left side $\mathcalA_m(H)$ and right side $\binom Vs$, where $T$ is adjacent to all its $s$-subsets. Replace each right vertex by $L$ copies with the same neighborhood. For any left-side family $\mathcalA$, its neighborhood in this new graph has cardinality $L|\partial_s\mathcalA|\geq|\mathcalA|$ by Lemma~\ref{lem:sampling}. Hall's marriage theorem~\cite{Hall} gives a matching saturating the left side. Assign each tuple to the original right vertex corresponding to its matched copy.
\end{proof}

\begin{theorem}[Shadow criterion]\label{thm:criterion}
Let $1\leq s<t$ be fixed integers and let $D,a$ be fixed positive constants, with $D$ an integer. Suppose a finite hypergraph $H$ has VC-dimension at most $D$ and satisfies the $(s,t,a)$-shadow condition. Then there is a constant $C=C(D,s,t,a)$ with the following property. For every $k\geq2$ and integer $m\geq C k^{1/(t-s)}$, there is a $k$-coloring of $\binom Vt$ that is polychromatic on every hyperedge of cardinality exactly $m$. If each hyperedge of size at least $m$ contains an $m$-element hyperedge, the same coloring is $(t,k,m)$-polychromatic.
\end{theorem}
\begin{proof}
Write $r=t-s$ and assume $m\geq2t$. Let $\alpha$ be the assignment from Lemma~\ref{lem:assignment}. For every $s$-subset $S\subseteq V$, define the family of completions
\[
 \QQ_S=\{T\setminus S:T\in\mathcalA_m(H),\ \alpha(T)=S\}.
\]
Each member of $\QQ_S$ is an $r$-tuple disjoint from $S$. The assignment gives
\begin{equation}\label{eq:capacity}
 |\QQ_S|\leq \lceil\mathrm e\,a\,m^r\rceil\leq (\mathrm e\,a+1)m^r.
\end{equation}
Unlike in the geometric proof, the family $\QQ_S$ of completions need not contain all $r$-tuples on a set of vertices. We therefore color the completions themselves. Equip $\QQ_S$ with the \emph{containment hypergraph} having hyperedges $\QQ_S(e)=\{U\in\QQ_S:U\subseteq e\}$ for $e\in E$. By a result of Alon et al.~\cite[Proposition~7]{AJKSY}, the hypergraph on $\binom Vr$ with hyperedges $\binom er$, $e\in E$, has VC-dimension $O(D\log(r+1))$. The containment hypergraph on $\QQ_S$ is a trace of this hypergraph, so the same bound holds for it.

Now fix an $m$-element hyperedge $e$. Every $t$-tuple in $e$ is $m$-shallow, and is assigned to one of its $s$-subsets. Hence
\[
 \sum_{S\in\binom es}|\QQ_S(e)|=\binom mt.
\]
Some $S\in\binom es$ therefore satisfies
\begin{align}
 |\QQ_S(e)| &\geq \frac{\binom mt}{\binom ms}
 =\frac{s!}{t!}\prod_{j=s}^{t-1}(m-j)\notag\\
 &\geq \frac{s!}{2^r t!}m^r.\label{eq:local-large}
\end{align}
The final inequality uses $m\geq2t$.

Apply Lemma~\ref{lem:local} separately to each containment hypergraph on $\QQ_S$, with $M=m^r$, $A=\mathrm e\,a+1$ and $B=s!/(2^r t!)$. In view of~\eqref{eq:capacity}, if $k\leq c m^r$ for a suitable constant $c>0$, this gives a coloring $\psi_S:\QQ_S\to[k]$ such that every range of size at least $Bm^r$ has all $k$ colors. The constant $c$ is uniform over all $S$ and $H$.

For every $m$-shallow tuple $T$, define
\[
 \chi(T)=\psi_{\alpha(T)}(T\setminus\alpha(T)),
\]
and color the other tuples arbitrarily. Each tuple receives only one color. For an $m$-element hyperedge $e$, choose $S$ satisfying~\eqref{eq:local-large}. For every $i\in[k]$, the set $\QQ_S(e)$ contains a completion $U$ with $\psi_S(U)=i$. Since $S\subseteq e$ and $U\subseteq e$, the tuple $S\cup U$ is contained in $e$ and has color $i$. This proves the claim for all $m$-element hyperedges simultaneously. Take $C$ large enough that $m\geq2t$ and $k\leq cm^r$ whenever $m\geq Ck^{1/r}$. The last assertion follows by applying the result to an $m$-element subhyperedge of any larger hyperedge.
\end{proof}

\paragraph{The shadow condition for pseudo-discs.}
Theorem~\ref{thm:criterion} also recovers the geometric bound of Theorem~\ref{thm:pseudodisks}. Indeed, let $\BB$ be a family of exact $t$-element hyperedges of a trace of a pseudo-disc hypergraph. Put $U=\bigcup_{T\in\BB}T$. Every member of $\BB$ is still an exact $t$-element hyperedge on $U$, and hence induces a clique in the graph $G_t$ on $U$. As in Section~\ref{sec:disks}, this graph is $\lfloor6\mathrm e\,t\rfloor$-degenerate. Each tuple in $\BB$ is determined by its first vertex and $t-1$ of its later neighbors. Therefore
\[
 |\BB|\leq |U|\binom{\lfloor6\mathrm e\,t\rfloor}{t-1}
 =\binom{\lfloor6\mathrm e\,t\rfloor}{t-1}|\partial_1\BB|.
\]
Thus the $(1,t,a_t)$-shadow condition holds for $a_t=\binom{\lfloor6\mathrm e\,t\rfloor}{t-1}$. Together with bounded VC-dimension and the enlargement to $\RR'$ in Section~\ref{sec:disks}, the shadow criterion implies the same bound. The direct geometric proof avoids the matching argument altogether.

\paragraph{Hypergraphs of bounded VC-dimension.}
It remains to show that the shadow condition holds for hypergraphs of bounded VC-dimension. We use the standard down-compression argument behind the Sauer--Shelah lemma~\cite{Sauer,Shelah}. The following lemma gives a bound in terms of the lower shadow of an arbitrary subfamily, rather than in terms of the size of the entire vertex set.

\begin{lemma}[VC-dimension and shadows]\label{lem:compression}
Let $d\geq1$ and $t>d$. If $\BB\subseteq\binom Vt$ has VC-dimension at most $d$, then
\begin{equation}\label{eq:compression}
 |\BB|\leq 2^d|\partial_d\BB|.
\end{equation}
\end{lemma}
\begin{proof}
For a family $\mathcal C\subseteq2^V$ and a vertex $v\in V$, perform the following down-compression: replace each $C\in\mathcal C$ containing $v$ by $C\setminus\{v\}$ if the latter set is not already in $\mathcal C$; keep all other sets. This preserves the number of sets. It does not increase VC-dimension: if the compressed family shatters $X$ not containing $v$, the original family has the same traces on $X$; if $v\in X$, each trace $Y\subseteq X$ containing $v$ must be realized by an original member containing $v$ whose deletion of $v$ was also in the original family. The original family thus realizes both $Y$ and $Y\setminus\{v\}$ for every such $Y$, and shatters $X$ as well.

Repeat nontrivial down-compressions starting from $\BB$. This process terminates because the sum of the cardinalities of the members strictly decreases at every nontrivial step. The resulting family $\BB'$ is downward closed, has $|\BB'|=|\BB|$, and has VC-dimension at most $d$. Every set in $\BB'$ therefore has size at most $d$ (a member of a downward-closed family is shattered). Moreover each member of $\BB'$ is contained in some original member of $\BB$, since the compressions only delete vertices. Since every original member has size $t>d$, each member of $\BB'$ is contained in a member of $\partial_d\BB$. Each $d$-set has exactly $2^d$ subsets, and consequently
\[
 |\BB|=|\BB'|\leq\left|\bigcup_{S\in\partial_d\BB}2^S\right|
 \leq2^d|\partial_d\BB|.
\]
\end{proof}

\begin{proof}[Proof of Theorem~\ref{thm:vc}]
Let $H$ have VC-dimension at most $d$. Every subfamily of every trace $H[X]$ also has VC-dimension at most $d$. Thus Lemma~\ref{lem:compression} verifies the $(d,t,2^d)$-shadow condition for $H$. Apply Theorem~\ref{thm:criterion} with $s=D=d$ and $a=2^d$, and use shrinkability to extend the coloring from $m$-element hyperedges to every hyperedge of size at least $m$. With $m=\lceil C_{d,t} k^{1/(t-d)}\rceil$, this gives the claimed bound.
\end{proof}

\section{Discussion and open problems}
\label{sec:conclusion}

\paragraph{Packing disjoint tuple nets.}
Recall that an $\eps$-$t$-net for a hypergraph $H=(V,E)$ with $n=|V|$ is a family of $t$-tuples such that every hyperedge of size at least $\eps n$ contains one of those tuples~\cite{AJKSY}. A polychromatic tuple coloring partitions $\binom Vt$ into pairwise disjoint tuple nets whenever its threshold is at most $\lceil\eps n\rceil$. Thus our results imply the following.

\begin{corollary}\label{cor:nets}
Let $0<\eps<1$, $n=|V|$, and suppose $\eps n$ is sufficiently large in terms of the fixed parameters.
\begin{enumerate}
 \item If $H$ is induced by planar points and pseudo-discs and $t\geq2$ is fixed, then $\binom Vt$ can be partitioned into $\Omega_t((\eps n)^{t-1})$ disjoint $\eps$-$t$-nets.
 \item If $H$ is shrinkable, $\vc(H)\leq d$, and $t>d$ is fixed, then $\binom Vt$ can be partitioned into $\Omega_{d,t}((\eps n)^{t-d})$ disjoint $\eps$-$t$-nets.
\end{enumerate}
\end{corollary}
\begin{proof}
Let $r=t-1$ in the first case and $r=t-d$ in the second. Both theorems give a threshold at most $\lceil Ck^{1/r}\rceil$ with the appropriate constant $C$. Take $k=\lfloor(\eps n/C)^r\rfloor$, which is at least two for sufficiently large $\eps n$. Then the threshold is at most $\lceil\eps n\rceil$, and every color class is an $\eps$-$t$-net. The classes are disjoint and partition all tuples. If there is no hyperedge of size at least $\eps n$, one may instead take any partition into $k$ nonempty classes; this is possible for sufficiently large $\eps n$ (enlarge $C$ if needed).
\end{proof}

If at least one hyperedge has size $\lceil\eps n\rceil$, the number of pairwise disjoint $\eps$-$t$-nets is at most $\binom{\lceil\eps n\rceil}{t}$. The preceding corollary is therefore a polynomial lower bound on the number of disjoint nets, not in general a tight estimate. It complements the packing result of Biniaz et al.~\cite{BDMSSS,BDMSSSfull} for hypergraphs having linear polychromatic vertex-coloring thresholds, which need not hold for discs.

\paragraph{Constructive aspects.}
For an explicitly listed hypergraph and fixed parameters, the proofs give polynomial-time constructions. In the geometric setting, compute the graph of shallow pairs and a degeneracy ordering, and obtain tuple colors from the local vertex colorings; in the abstract setting, enumerate shallow tuples and compute their assignment by capacitated bipartite matching. The constant-size local nets can be found by exhaustive search on each local ground set. For pseudo-disc hypergraphs, this statement concerns the enlarged trace family used in the proof, when it is given explicitly. No sharp geometric running time is claimed.

\paragraph{Open problems.}
For discs, can the linear threshold for pairs be improved to $O(\sqrt{k\log k})$, or to the counting lower bound $\Theta(\sqrt{k})$? Our degeneracy argument assigns pairs to their first endpoint. Although each later neighborhood admits $\Omega(m^2)$ local colors on its internal pairs, these local colorings overlap and need not agree. Obtaining globally compatible colorings remains open. For $t\geq3$, the corresponding question is whether the pseudo-disc exponent $1/(t-1)$ can be replaced by $1/t$.

For balls in $\mathbb R^D$, the VC-dimension bound gives polynomial thresholds for $t>D+1$. Biniaz et al.~\cite{BDMSSS,BDMSSSfull} obtained exponential bounds already for $t=\lfloor(D+3)/2\rfloor$. It is open here whether the lower-shadow criterion applies at that smaller tuple size in higher dimensions; a bound relative to the shadow of every subfamily of exact $t$-element traces would suffice.

\bibliographystyle{plainnat}
\bibliography{references}

\section*{AI Statement}
The central ideas underlying the results in this paper originated with the authors. We used ChatGPT-5.6 Sol to assist in developing these ideas and preparing an initial draft. The manuscript was subsequently rewritten by the authors into its present form.
\end{document}